 \documentclass[aps,twocolumn,showpacs]{revtex4}
 \usepackage{graphics}
 
\begin{document}
 \bibliographystyle{apsrev}
 \def\half{{1\over 2}}
 \def \D {\mbox{D}}
 \def\curl {\mbox{curl}\,}
 \def \ep {\varepsilon}
 \def \lleq {\lower0.9ex\hbox{ $\buildrel < \over \sim$} ~}
 \def \ggeq {\lower0.9ex\hbox{ $\buildrel > \over \sim$} ~}
 \def\beq{\begin{equation}}
 \def\eeq{\end{equation}}
 \def\ber{\begin{eqnarray}}
 \def\eer{\end{eqnarray}}
 \def \apl {ApJ, }
 \def \aps {ApJS, }
 \def \pd {Phys. Rev. D, }
 \def \prl {Phys. Rev. Lett., }
 \def \pl {Phys. Lett., }
 \def \np {Nucl. Phys., }
 \def \l {\Lambda}

  \def\apj{{Astroph.\@ J.\ }}
  \def\mn{{Mon.\@ Not.\@ Roy.\@ Ast.\@ Soc.\ }}
  \def\asta{{Astron.\@ Astrophys.\ }}
  \def\aj{{Astron.\@ J.\ }}
  \def\prl{{Phys.\@ Rev.\@ Lett.\ }}
  \def\pd{{Phys.\@ Rev.\@ D\ }}
  \def\nucp{{Nucl.\@ Phys.\ }}
  \def\nat{{Nature\ }}
  \def\plb {{Phys.\@ Lett.\@ B\ }}
  \def \jetpl {JETP Lett.\ }

\title{Submillimeter corrections to gravity and the
metastability of \\ white dwarf and neutron stars}
\author{Mofazzal Azam$^1$ and \ M. Sami$^2$}
\address{$^1$Theoretical Physics Division, Bhabha Atomic Research Centre\\
Mumbai, India\\
$^2$Department of physics, Jamia Millia Islamia, New Delhi.}
\begin{abstract}
The string inspired higher dimensional theories
suggest modification of Newton's law at submillimeter
length scales. Inter-particle distances in white dwarf and neutron
stars are $10^{-10}~cms$ and $10^{-13} cms$ respectively, and
therefore, the effects of of short distance corrections
to gravity deserve investigation.
We show, by carrying out explicit analytical many-body
calculations that, in the presence of corrections, the normal state
of these compact stars become metastable. The actual quantum mechanical
ground state of the stars turns out to be unstable.
However, the tunneling  probability to the unstable ground state
is so small that the stars may remain trapped in the metastable
state practically for ever.
\end{abstract}
\maketitle
\vskip .8 in
\section{INTRODUCTION}
The brane-world scenario proposed by  Randall and Sundrum \cite{ran,othors},
gives rise to modification of gravity
at submillimeter distance scales; the corrections are in higher powers of inverse distance
and appear as additional terms in the Newtonian potential.
The modified potential is given by,
\begin{eqnarray}
\Phi(r)= -\frac{GM}{r}(1+\frac{2l^2}{3r^2})
\end{eqnarray}
where $l$ is the curvature scale of 5-dimensional anti-deSitter space time,
and as before,
$G$ is the usual Newton's constant of gravity, $M$ is the mass, $r$ is the
distance in 3-space. In the setting above, $l$ has the dimension of length
in 3-space. Henceforth, we will take $l_{s}^{2}=\frac{2}{3}l^2$.\par
The astrophysical consequences of
short distance modifications of gravity are discussed in Refs.\cite{roy,wise}.
The average inter-particle
distance in compact stars such as the
white dwarf and neutron stars are $10^{-10}~cms$ and
$10^{-13}~cms$ respectively, and
therefore, the submillimeter
corrections, in principle, would effect such parameters as mass and radius,
and also the mass-radius relationship of the stars.
Wiseman \cite{wise} has carried
out a detailed numerical study of the neutron stars on the brane and
has claimed that these parameters of the neutron stars
are not effected by brane world corrections.\par
In Ref.\cite{azam}, we have carried out
some simple analytical many-body calculations using the brane
modified gravitational potential.
Our approach was based on many-body theory in
semi-relativistic quantum mechanics.
It was demonstrated that the inclusion
of the brane correction terms does not effect the Chandrasekhar
ground state which is consistent with the
results obtained by Wiseman using numerical technique \cite{wise}.
In this paper, we intend to show that, although correction
terms do not effect the variational minimum of the energy
corresponding to Chandrasekhar ground state, they give
rise to two new features of energy as a function of the radius:
a positive maximum very close to the origin and
a new unbounded minimum at the origin itself. The unbounded
minimum indicates the absence of true quantum mechanical
ground state and, there by, a quantum mechanical instability.
This implies that the normal Chandrasekhar ground state of
these compact stars become metastable.
We show that the tunneling
probability of the stars from the metastable state to the
unstable ground state is vanishingly small.
\section{The kinetic energy of relativistic degenerate electron gas}
We consider quantum mechanics of a self-gravitating system of
large number of fermions  in a compact spherical volume, $V$,
with a large constant density
such that fermions are relativistic. We will, first, consider here
the white dwarf stars and, then, argue that our results
can also be carried over to neutron stars.
Consider a white dwarf star whose constituents
are atoms of mass number $A$ and atomic number $Z$. Let the number
of atoms in the star be $N_A$. The star is charge neutral,
and therefore, the total number of electrons in star, $N_e=Z N_A$
and the total number of nucleons (protons$+$neutrons) $N_n=A N_A$.
Latter, we will see that, $N_e/N_n=Z/A=Y_e$ is an important
characteristic of a white dwarf star. Our basic assumption are:\\
(1) For white dwarf stars general relativistic effects can be
neglected.\\
(1) Atomic nuclei are practically motionless and
provide the entire gravitational attraction.\\
(2) The star is so  dense that electrons constitute a degenerate
relativistic Fermi gas at zero temperature.\\
The kinetic energy of the star
is the energy of the relativistic degenerate free Fermi
gas which can be evaluated as follows\cite{huang}.
The single-particle electron states are specified
by the momentum ${\bf p}$ and the spin quantum numbers $s=\pm 1/2$.
The single-particle energy levels are independent of $s$ :
\begin{eqnarray}
\epsilon_{{\bf p}s}=\sqrt{({\bf p}c)^2 +(m_ec^2)^2}-m_ec^2  \nonumber
\end{eqnarray}
where $m_e$ is the electron mass. The energy of
free Fermi electron gas is given by,
\begin{eqnarray}
{\mathcal{E}}_F= {\mathcal{E}}_0 \bar{{\mathcal{E}}}(x_F)-N_em_ec^2~;~~
{\mathcal{E}}_0 =\frac{(m_e c^2)^4}{(c\hbar)^3}V;~~~~~~~~~~~~~~~~~~
\nonumber\\
\bar{{\mathcal{E}}}(x_F)=\frac{1}{8\pi^2}\Big{[}x_F(1+2x_{F}^{2})
\sqrt{1+x_{F}^{2}}-ln(x_F + \sqrt{1+x_{F}^{2}})\Big{]} \nonumber
\end{eqnarray}
where $x_F=p_F/m_e c$ and the Fermi momentum is given by
$ p_F=\hbar(3\pi^2 n_e)^{1/3} ~(n_e =N_e/V)$.

In terms of $x_F$, ${\mathcal{E}}_0=3\pi^2N_em_ec^2/x_{F}^3$.
Therefore, the kinetic energy of the electron gas
for constant electron density is given by
\begin{eqnarray}
K.E.\equiv {\mathcal{E}}_F=\frac{3\pi^2N_em_ec^2}{x_{F}^3}
\bar{{\mathcal{E}}}(x_F)-N_em_ec^2
\end{eqnarray}
For
variable electron density, a similar expression has been
derived by Daubechies in a very general and
mathematically rigorous way (\cite{dau}, see also \cite{lieb}).
We will mainly be considering spherically symmetric stars
with a constant density profile.
Following reference \cite{jack}, we express
the dimensionless Fermi momentum, $x_F$,
in terms of mass and radius of the white dwarf stars as follows.
Let, $M_0$ and $R_0$, of dimensions of
mass and length be given by,
\begin{eqnarray}
M_0=\frac{5}{6}\sqrt{15\pi}\Big{(}~\frac{c\hbar}{Gm_{n}^{2}}~\Big{)}^{3/2}
m_nY_{e}^{2}=10.599~M_{\odot}Y_{e}^{2} \nonumber\\
 R_0=\frac{\sqrt{15\pi}}{2}
\Big{(}~\frac{c\hbar}{Gm_{n}^{2}}~\Big{)}^{1/2}
\frac{c\hbar}{m_ec^2}Y_e=17250 km \times Y_e \nonumber\\
\end{eqnarray}
For helium white dwarf stars, $Y_e=1/2$, therefore,
$M_0=2.650 M_{\odot}$ and $R_0=8623 km$.\par
It will also be convenient to introduce the dimensionless mass and radius of stars
as $\bar{M}\equiv M/M_0$ and $\bar{R}\equiv R/R_0$ in terms of which
$x_{F}={\bar{M}^{1/3}}/{\bar{R}}$.
\section{The interaction potentials and the potential energy}
We consider a spherically symmetric star of mass $M$ and radius
$R$, with a constant mass density profile $\rho_A(r)= \rho=constant$
for $0\leq r\leq R$, and $\rho_A(r)= 0$ for $r>R$.
Therefore, $M=N_Am_A=\frac{4\pi}{3}\rho R^3$.
In what follows we briefly describe the potential energy of the system
in presence of short range corrections due to
Randall-Sundrum (RS) and Yukawa corrections (see Ref.\cite{azam} for details)
At any point $\vec{r}$ inside a solid sphere, the potential
consists of two parts:
one part is due to the inner sphere bellow the point $\vec{r}$ and second
part is due to the outer spherical shell.
Let $\Phi^{(1)}(r,a)$ be the potential outside a solid
sphere of radius $a$, and $\Phi^{(2)}(r,b,R)$ be the potential inside
a spherical shell of inner radius $b$ and outer radius $R$. The
potential energy is given by,
\begin{eqnarray}
U=\frac{1}{2}\Big{[}\int_0^R 4\pi\rho r^2dr
\Big{(}\Phi^{(1)}(r-\epsilon,r)+\Phi^{(2)}(r,r+\epsilon,R)\Big{)}
\Big{]}\nonumber\\
\end{eqnarray}
At the end of the calculations, we try to take the limit
$\epsilon\rightarrow 0$. If the limit does not exist, we take
$\epsilon$ to be the smallest inter-particle distance.\\
Using the expressions for the potentials $\Phi^{(1)}_{RS}(r,a)$
and $\Phi^{(2)}_{RS}(r,b,R)$ obtained in Ref.\cite{azam},
we obtain the potential energy corresponding to RS corrections in Eq. (1),
\begin{eqnarray}
U_{RS}=-\frac{3}{2}\frac{GM^2l_{s}^{2}}{R^3}ln\frac{R}{\epsilon}
\end{eqnarray}
Here $\epsilon$ is the average inter-particle distance in the star.
For a system of Fermions, it will be proportional to $R/N^{1/3}$.
The precise numerical factor which is of order one, will not matter
for our discussions, and therefore, we will simply take
$\epsilon=R/N_{n}^{1/3}$. The potential energy can, therefore,
be written as
\begin{eqnarray}
U_{RS}=-\frac{3}{2}\frac{GM^2l_{s}^{2}}{R^3}ln(N_{n}^{1/3})
\end{eqnarray}
In terms of the variables $M_0$ and $R_0$,
\begin{eqnarray}
U_{RS}=-\Bigg(\frac{5l_{s}^{2}ln(N_{n}^{1/3})}{2R_{0}^{2}}\Bigg)
\Bigg(\frac{3}{5}\frac{GM^2}{R}\Bigg)\Bigg(\frac{R_{0}^{2}}{R^2}\Bigg)
\nonumber
\end{eqnarray}
Let us call the terms in the first bracket $\beta$.
In terms of the Fermi momentum, $x_F$ the total potential energy
corresponding brane modified potential given by Eq. (1) is
\begin{eqnarray}
U=-N_em_ec^2\bar{M}^{2/3}x_F-N_em_ec^2\beta x_{F}^{3}
\end{eqnarray}
where the first part in the above equation is contributed
by the Newtonian potential.
For submillimeter corrections, $\beta\simeq 10^{-19}$ for helium stars.
For other stars it varies within a order magnitude of this number.\\

Yukawa interaction is a very generic
short range interaction (see \cite{azam1} for details)
and therefore, the results obtained for this potential is
expected to hold qualitatively for most short range potentials.
On phenomenological grounds we require, that the
strength of the potential
due to the correction term at the millimeter
length scale, at the best, be comparable to gravity. This
means that the parameter, $\alpha$, characterizing the strength
of potential, be a number of the order of unity.
We will keep in mind this fact while preserving $\alpha$
in all our calculations \cite{bro}.

For spherically symmetric star of radius
$R$ and constant mass density $\rho$, the many-body potential
at any point $r~$, $0\leq r\leq R$ is given by \cite{azam1},
\begin{eqnarray}
\Phi_{Y}(r)=-\frac{4\pi G\alpha \rho}{\mu^2}
\Bigg{(}~1-\Big[\Big{(}R+\frac{1}{\mu}\Big{)}
e^{-\mu R}\frac{sinh(\mu r)}{r}\Big]~\Bigg{)}\nonumber\\
\end{eqnarray}
Let us note that the second term in the bracket is very small
and is negligible
in the body of the star but it is large close to the boundary
and will give rise to surface tension as in the case of
Randall-Sundrum corrections. We are
interested in the balance of pressure and gravity (along with various
corrections) in the body of the star, and therefore,
we neglect this term.
Thus in the potential given in Eq.(8), the only
relevant part is:
$\Phi_{Y}(r)=-4\pi G\alpha \rho/\mu^2$.
Therefore, the potential energy,
\begin{eqnarray}
U_Y=\int_{0}^{R}4\pi r^2\rho\Phi_{Y}(r)dr
=-N_em_ec^2\gamma x_{F}^3
\end{eqnarray}
where $\gamma=\Big(\frac{5\alpha}{\mu^2 R_{0}^{2}}\Big)$.
For submillimeter corrections $\gamma\simeq 10^{-19}$ for helium stars.
Note that the form of potential energy for both RS and
Yukawa corrections are cubic in $x_F$.
The dimensionless parameters $\beta$ (for RS)  and
$\gamma$ (for Yukawa) are of same order of magnitude and extremely
small. Henceforth, we use $\beta$ for both.\\
Now, the relativistic quantum mechanical total energy is,
$E(\bar{M},x_F)=N_em_ec^2\Xi(\bar{M},x_F)$, where
\begin{eqnarray}
\Xi(\bar{M},x_F) =\Bigg[\frac{3\pi^2\bar{{\mathcal{E}}}(x_F) }
{x_{F}^3} -1
-\bar{M}^{2/3}x_F -\beta x_{F}^{3}\Bigg]
\end{eqnarray}
In Eq.(10), the first two terms arise from the kinetic
energy given in Eq.(2), the third term is due to the Newtonian
potential energy (see reference \cite{jack}), and the last term
comes, as shown in Eqs.(7) $\&$ (9), from potential energy
due to the short range corrections.
In the absence of correction terms, the total energy is
$E_c=N_em_ec^2\Xi_c(\bar{M},x_F)$, where
\begin{eqnarray}
\Xi_c(\bar{M},x_F)=\Bigg[\frac{3\pi^2\bar{{\mathcal{E}}}(x_F) }
{x_{F}^3} -1-\bar{M}^{2/3}x_F\Bigg]
\end{eqnarray}

\section{Ground state of the compact stars in the presence of short
range correction to gravity}
\subsection{The white dwarf stars}
Large quantum systems such as white dwarf stars are
stable if the ground state energy per particle is bounded
bellow. For a fixed value of $\bar{M}$, we look for the extrema of
per-particle energy, $\Xi=E/(N_em_ec^2)$,  as a function of $x_F$.
In the absence of corrections, it is known from the
works of Chandrasekhar (\cite{chandra}, see also \cite{lieb})
that bellow some critical mass, $\bar{M}_c$, the energy
functional, $\Xi_c$ given in Eq.(11),
has a unique minimum. As we change the mass, the location
of the minimum changes. By definition,
$x_F$ is inversely proportional to the radius, and therefore,
the variation of the location of the minimum with the mass
gives the mass-radius relationship of the star.\\
We choose $\bar{M}=\bar{M}_1(\simeq 0.25)$,
such that the star is Chandrasekhar
stable with energy minimum at $x_F=\bar{x}_1(\simeq 0.8)$.
Now consider the energy with correction terms, Eq.(7),
for $\bar{M}=\bar{M}_1$. The correction term is too small,
and become relevant only for very very large values of $x_F$
and therefore, the local minimum, which exist for small $x_F$
in the absence of corrections, will persist in the presence of the
correction terms.
Therefore, as before, energy as function of $x_F$
will have a local minimum at
$x_F=\bar{x}_1$.  However, for very large value of $x_F$, the energy
approaches $-\infty$. To see this let us note, that for $x_F>>1$,
$\frac{3\pi^2{\mathcal{E}}(x_F)}{x_{F}^{3}}=\frac{3}{4}x_F$.
Therefore,
\begin{eqnarray}
\Xi=\Bigg[\frac{3}{4}-\bar{M}_{1}^{2/3}\Bigg]x_F-1-\beta x_{F}^3
\end{eqnarray}
Since $\bar{M}_{1}<M_c$, the expression in the square bracket
in eq.(12) is positive.
It is easy to see that $\Xi$, given above has a maximum
for $x_F=\sqrt{\frac{1}{2\beta}
\Big(\frac{3}{4}-\bar{M}_{1}^{2/3}\Big)}$.
The value of the function
$\Xi$ at the  point of the maximum is very large.
Beyond the maximum, $\Xi$ rolls
down to $-\infty$.
To summarize, the energy with correction term has the
following structure: for small $x_F$, there is a minimum, then for very
large value there is a maximum after which it approaches $-\infty$.
The dimensionless Fermi momentum, $x_F$,
is inversely proportional to the radius, R.
Therefore, energy as a function of $R$ has a minimum for some
finite value of $R$, say $R=R_1$
corresponding to the Chandrasekhar state.
In addition it has a maximum very close to $R=0$ (say, for $R=R_2$),
followed by an unbounded minimum at the origin.
This makes the Chandrasekhar state metastable.
In the dimensionless unit that we introduced, $R_1$ is of
order one, where as $R_2\sim \sqrt{\beta}\sim 10^{-9.5}$, therefore,
$R_1/R_2>>1$.\\
The potential and force due to the short range
correction terms are similar to
those in nuclear physics except that the strength is much
weaker and the range is much larger. Therefore, a reference
to nuclear physics may unveil qualitatively
the relevant physical features.
Consider, the energy-per-particle without the Newtonian term,
$E'(\bar{M},x_F)=N_em_ec^2\Xi'(\bar{M},x_F)$, where
\begin{eqnarray}
\Xi'(\bar{M},x_F)  =\Bigg[\frac{3\pi^2\bar{{\mathcal{E}}}(x_F) }
{x_{F}^3} -1-\beta x_{F}^{3}\Bigg]
\end{eqnarray}
This function has a maximum and an unbounded minimum.
There is no other minimum. This feature is reminiscent
of heavy nuclei, with nucleons interacting via the Yukawa
potential. With suitable interpretation
of the parameters,
Eq.(13), can, in fact, be carried
over to nuclear physics for
quantitative treatment of  heavy nuclei. It is well known
in nuclear physics that the zero-point kinetic energy
whether relativistic or non-relativistic can not
balance the attractive potential energy of the Yukawa
interaction. The problem is solved by invoking a very short range
hard core repulsive potential which provides a lower bound and
removes the unbounded minimum. Without this hard core repulsive
potential, heavy nuclei can not be stable.\\
In our case we do not have a repulsive hard core potential.
But the inclusion
of long range Newtonian potential in Eq.(13)
makes it identical to  Eq.(10), and therefore,
creates, bellow a critical mass, a local minimum
in the energy function without effecting the other two features:
a maximum and an unbounded minimum.
The original Chandrasekhar ground state shows up as a metastable
state in the local minimum of the energy functional.
\subsection{The neutron stars}
The neutron stars can be analyzed exactly in the same way
as the white dwarf stars except for the fact that for neutron
stars the general relativistic effects would play some role.
In what follows, we, first, neglect these effects and
comment on them at the end of this section.
The first thing that should be noted is that
both the kinetic and the potential energy in the neutron stars
are due to the neutrons alone.
Therefore, all the formulae and equations which are valid for the
white dwarf stars are also valid for the neutron with the following
replacements: (electron density, $n_e~ \rightarrow$ neutron density,
$n_n$), (electron mass $m_e~ \rightarrow$ neutron mass, $m_n$), (number
of electrons, $N_e~ \rightarrow$ number of neutrons $N_n$), and
($Y_e \rightarrow 1$). This leads to the change in the numerical value
of the following quantity, $M_0\rightarrow M_0^{(n)}=10.599~M_{\odot}~$,
$R_0\rightarrow R_0^{(n)}= 9.40 km~$,$\beta\rightarrow \beta^{(n)}=
\simeq 10^{-13}~$, $\gamma\rightarrow \gamma^{(n)}=\simeq 10^{-13}$.
The ground state of the neutron stars can be found just
as in the case of white dwarfs. We reach the similar conclusion
that, in the presence of short range corrections,
neutron stars are metastable.

The general relativistic effects in the compact neutron stars ,
in our opinion, influences the Chandrasekhar ground state
and the values of parameters such as mass and radius
that are associated with it. It also changes the critical mass for
the stability of the star.
However, it should be clear from the discussions in the earlier sections
that the short range correction to Newton's law of gravity does not effect
the critical mass for the stability of the stars given by
Chandrasekhar limit. It also does not effect
the mass, radius or mass-relationship when the star
is in the metastable Chandrasekhar ground state. This suggests that
our results, in all likely hood, are valid in the presence of
general relativistic effects.
\subsection{The probability of tunneling to the unstable ground state}
Let us provide an estimate of the probability of tunneling from
the metastable state to the unstable ground state. It is obvious
that the peak, at the maximum of the energy functional, that
separates the metastable state (local minimum) from the
unstable ground state (global minimum), acts as an energy
barrier for the tunneling process. Let us recall some of
the formulae from quantum mechanics on barrier penetration.
Suppose
that potential between two points, $x_1$ and $x_2$ increases
monotonically, the point, $x_2$, is the peak of the potential
and the energy of the particle crossing the barrier equals
to its' value at the lower end. Then, the transmission
coefficient or the tunneling probability,
$T\sim e^{-S/\hbar}$, where the action,
$S=\int_{x_1}^{x_2} p.dx$, and $S/\hbar>>1$.
For the compact stars, let us represent by $R_1$, the radius of the
star in the metastable ground state and, by $R_2$, the radius
of the star at the local maximum
separating the unstable ground state. As argued in the earlier
sections, $R_1>>R_2$.
The typical momenta of individual particle in these states are,
approximately, given by the Fermi momenta, $\hbar N^{1/3}/R_1$ and
$\hbar N^{1/3}/R_2$.
Tunneling of the compact stars through the energy barrier at the
maximum of the energy functional amounts to the change in the
radius of spheres, occupied by individual
particles, from $R_1/N^{1/3}$ to $R_2/N^{1/3}$.
This means that inter-particle
distances change from $R_1/N^{1/3}$ to $R_2/N^{1/3}$, and
as a result, momenta of individual particles change from
$\hbar N^{1/3}/R_1$ to $\hbar N^{1/3}/R_2$.
We will first calculate probability
for such a tunneling for a single particle. The action $S$, for this
can be given by $S=\int_{x_1}^{x_2} p.dx\approx \Delta p \Delta x=
(\hbar N^{1/3}/R_2-\hbar N^{1/3}/R_1)(R_1/N^{1/3} - R_2/N^{1/3})$.
The condition, $R_1>>R_2, \rightarrow S/\hbar=R_1/R_2$.
Therefore, the tunneling probability for a particle,
$T_{single}\sim e^{-S/\hbar}\simeq e^{-R_1/R_2}$. When all the
particle, simultaneously, undergo such transitions,
the compact star will tunnel through the energy barrier.
For a star with particle number, $N$, the tunneling probability
is, $T\approx e^{-N(R_1/R_2)}$. For  typical
white dwarf and neutron stars, particle number,
$N\sim 10^{57}$, which is a huge number, and therefore, the
tunneling probability to the unstable ground state is
extremely small, and therefore, negligible.
\section{conclusions}
We have analysed in details the effects of short range corrections to
gravity on the compact stellar objects such as
the white dwarf and neutron stars.
The corrections lead to two new effects. One of them is the
emergence of surface tension which is the characteristic of short
range interactions and has been discussed in detail in our previous
publication \cite{azam}. The second effect is that
the compact stars become metastable. We have shown that the tunneling probability
to the unstable ground state is so small that the stars once trapped
in the metastable state would remain there.
\section{ACKNOWLEDGEMENTS}
We thank Sergei Odintsov for useful comments.


\end{document}